**Tunable Non-local Spin Control in a Coupled Quantum Dot System**

N. J. Craig, J. M. Taylor, E. A. Lester, C. M. Marcus

*Department of Physics, Harvard University, Cambridge, Massachusetts 02138, USA*

M. P. Hanson, and A. C. Gossard

*Materials Department, University of California, Santa Barbara, Santa Barbara, California 93106, USA*

The effective interaction between magnetic impurities in metals that can lead to various magnetic ground states often competes with a tendency for electrons near impurities to screen the local moment (Kondo effect). The simplest system exhibiting the richness of this competition, the two-impurity Kondo system, is here realized experimentally in the form of two quantum dots coupled through an open conducting region. We demonstrate non-local spin control by suppressing and splitting Kondo resonances in one quantum dot by changing electron number and coupling of the other dot. Results suggest an approach to non-local spin control relevant to quantum information processing.

Gate-confined quantum dots have emerged as important systems for the study of the Kondo effect, a subtle many-electron effect in which conduction electrons in the vicinity of a spin impurity screen the spin to form a collective, entangled ground state at low temperatures (1). The connection between Kondo physics and quantum dots is most evident when an odd number of electrons confined within the dot act as a single spin coupled to electron reservoirs (2-5). Recently, molecule-like double quantum dots have also generated wide interest as controllable systems for studying exchange between coupled localized states (6-8) and as potential basic building blocks for quantum information processing, with proposed schemes for using double dots as sources of entangled electrons (9) and for two-qubit quantum gate operations (10).

It is known from bulk systems that magnetic impurities embedded in an electron sea interact with one another via an effective spin-spin interaction known as the Ruderman-Kittel-Kasuya-Yoshida (RKKY) interaction, mediated by conduction electrons (11-13). The RKKY interaction competes with local interactions between the impurity and conduction electrons that lead to the Kondo effect, and when dominant over Kondo interactions can give rise to complex bulk magnetic states such as spin glasses (14). Over the past two decades, the multiple-impurity Kondo system has proven to be a rich theoretical problem, exhibiting, among other features, a quantum phase transition between Kondo and RKKY regimes at a critical ratio of $(J/T_K)$ of order unity (depending on the particular geometry), where J is the RKKY interaction strength and $T_K$ is the single-impurity Kondo temperature (15-17). Recent theory has begun to extend the study of the two-impurity Kondo model to double quantum dots and related artificial spin systems (18-24). Experiments have explored the competition between the Kondo effect and exchange in directly coupled double quantum dots (7-8). However, a non-local RKKY-like interaction mediated by an interceding electron sea has not been described in an artificial system.

The device consists of two smaller peripheral quantum dots connected to a larger, open central dot, as shown in Fig. 1A (25). Measurements were made in a dilution refrigerator with a base electron temperature of ~85 mK, estimated from thermally broadened Coulomb blockade peaks measured on individual dots. Voltage bias spectroscopy on the left and right dots in the Coulomb blockade regime give Coulomb charging energies U ~ 800 μeV and level spacings Δ ~ 100 μeV for both dots. Differential conductances $dI/dV_{L(R)}$ of the left and right dots were measured simultaneously by applying voltage-bias excitations, V, consisting of dc, 11 Hz, and 27 Hz signals, to the open (bottom) lead of the center dot, and lock-in detecting ac currents at 11(27) Hz as well as dc currents at the left (right) reservoirs (both at virtual ground). Modeling the three-dot system as a voltage divider allows dc voltages $V_{L(R)}$ dropped across the left (right) dots to be readily extracted.

Setting the bottom point contact to one fully conducting spin-degenerate mode (conductance on the $2e^2/h$ plateau) configured the central dot to act as a confined but open conducting region coupling the two peripheral dots. Couplings of the left and right peripheral dots were set in the asymmetric Coulomb blockade regime, with relatively strong tunnel couplings $\Gamma_{L(R)}^{(c)}$ toward the central region and weak "outward" couplings $\Gamma_{L(R)}^{(l)}$ to the leads ($\Gamma_{L(R)}^{(l)} << \Gamma_{L(R)}^{(c)} \sim \Delta_{L(R)}$). This was done to ensure that any Kondo effect observed in the peripheral dots was associated with conduction electrons in the central dot, and not in the left and right leads. The left (right) dots were tuned to contain either an odd number, N (M), or even number, N±1 (M±1), of electrons by changing the voltage applied to gate $V_{gL}$ ($V_{gR}$).

Kondo effects in the asymmetric-coupling regime were investigated in the individual peripheral dots by pinching off the other peripheral dot entirely from the central region. Each dot individually showed characteristics of the Kondo effect, including elevated conductance through odd Coulomb blockade valleys, a zero-bias peak in the differential conductance $dI/dV_{L(R)}$ in odd valleys, and temperature dependence of valley height in qualitative agreement with theory. When either dot was in an even-occupancy valley, turning on its coupling to the central dot (set initially at zero, as described above) did not qualitatively affect signatures of the Kondo effect in the other dot.

A more interesting situation arises when both dots contained an odd number of electrons and the Kondo states in the peripheral dots interact. Figure 1B shows the relevant comparison: when the right dot contains an even number of electrons, the odd (N-electron) Coulomb blockade valley in the left dot exhibits Kondo signatures, including a pronounced zero-bias peak in $dI/dV_L$; however, when the right dot contains an odd number of electrons, the Kondo signatures in the left dot, including the zero-bias peak, are absent. Moving sequentially to the next even valley in the right dot brings back the Kondo signatures in the left dot. We interpret the suppression of Kondo signatures in the odd-odd case as indicating that RKKY interaction between the dots dominates the Kondo effect, forming either a overall spin-zero state (which has no Kondo effect) or a spin-one state with a much weaker Kondo effect at the temperatures measured.

Figure 2 illustrates the same effect with the roles of the dots reversed, measured in a different range of gate voltages. As anticipated, we now observe a suppression of the Kondo effect in the right dot—shown in the full plot of $dI/dV_R$ versus $V_R$ and $V_{gR}$—when the left dot contains an odd number of electrons (Figs. 2B,D). When the occupancy of the left dot is made even by the removal of one electron, the zero-bias signature of the Kondo effect in the right dot is recovered (Figs. 2A,C).

When both dots have an odd number of electrons, the suppression of the zero-bias peak in one dot can be tuned continuously by changing the central coupling strength of the other dot. Figure 3A shows the differential conductance $dI/dV_L$ of the left dot as the coupling $\Gamma_R^{(c)}$ of the right dot to the central region is tuned from strong to weak by changing the coupling gate voltage, $V_{gC}$. The zero-bias Kondo peak in the left dot first splits before being suppressed entirely as the right dot is coupled to the central dot. In contrast, when the right dot contains an even number of electrons, the strength of its coupling has little effect on the zero-bias peak of the left dot (Fig. 3B). The splitting of the zero-bias peak is a signature of quantum coherence between Kondo states on the peripheral dots (22, 26). The magnitude of the splitting corresponds to a splitting in source-drain voltage of $V_L \sim 0.12$ meV and does not depend strongly on the coupling of the right dot once it appears. This splitting is comparable to the width of the zero-bias peak in the left dot (FWHM $\sim 0.1$ mV, giving $T_K \sim 0.6$ K) before the right dot is coupled. However, it is not yet known if the similarity of scales for the Kondo peak width and the splitting is a general phenomenon. The physical mechanism that gives rise to the splitting and how its magnitude is related to the strength of the RKKY interaction is not a settled matter (26).

Both single-dot and coupled-dot configurations (not shown) show roughly linear peak splitting as a function of in-plane magnetic field in the range $B_\parallel \sim 2$–4T, with slopes of $\sim 70!\mu$eV/T. This slope is larger by a factor of $\sim 1.5$ than expected for the GaAs g-factor of 0.44, but is consistent with g-factor measurements in other devices made from the same wafer. Both the single-dot and coupled-dot cases show an unexpected strengthening of the zero-bias peaks with $B_\parallel$ before splitting is observed (for $B_\parallel < 2$T). This is not understood at present and will be investigated in more favorable device geometries in future work.

We have demonstrated coherent control of quantum dot spins by a non-local RKKY-like interaction. The present results suggest an approach to non-local control of spin and entanglement (27-29), which may be relevant to scaling of solid-state quantum information processing beyond the constraint of nearest-neighbor exchange.

## References

1.  An introductory review of the Kondo effect in quantum dots is: L. Kouwenhoven, L. Glazman, *Physics World* **14**, 33 (2001).

2.  L. I. Glazman, M. E. Raikh, *JETP Lett.* **47**, 452 (1988).

3.  T.K. Ng, P.A. Lee, *Phys. Rev. Lett.* **61**, 1768 (1988).

4.  Y. Meir, N.S. Wingreen, P.A. Lee, *Phys. Rev. Lett.* **70**, 2601 (1993).

5.  D. Goldhaber-Gordon *et al.*, *Nature* **391**, 156 (1998).

6.  A recent review of double quantum dots, with extensive citations to the literature may be found in W. G. van der Wiel et al., *Rev. Mod. Phys.* **75**, 1 (2003).

7.  H. Jeong, A. M. Chang, M. R. Melloch, *Science* **293**, 2221 (2001).

8.  J.C. Chen, A.M. Chang, M.R. Melloch, cond-mat/0305289 (2003).

9.  D. Loss, E. V. Sukhorukov, *Phys. Rev. Lett.* **84**, 1035 (2000).

10. D. Loss, D. P. DiVincenzo, *Phys. Rev. A* **57**, 120 (1998).

11. M. A. Ruderman, C. Kittel, *Phys. Rev.* **96**, 99 (1954).

12. T. Kasuya, *Prog. Theor. Phys.* **16**, 45 (1956).

13. K. Yosida, *Phys. Rev.* **106**, 893 (1957).

14. A. C. Hewson, *The Kondo Problem to Heavy Fermions*, Cambridge Studies in Magnetism (Cambridge University Press, Cambridge, 1993).

15. C. Jayaprakash, H. R. Krishnamurthy, J. W. Wilkins, *Phys. Rev. Lett.* **47**, 737 (1981).

16. B.A. Jones, C.M. Varma, *Phys. Rev. Lett.* **58**, 843 (1987).

17. B.A. Jones, C.M. Varma, *Phys. Rev. B* **40**, 324 (1989).

18. A. Georges, Y. Meir, *Phys. Rev. Lett.* **82**, 3508 (1999).

19. R. Aguado, D.C. Langreth, *Phys. Rev. Lett.* **85**, 1946 (2000).

20. W. Izumida, O. Sakai, *Phys. Rev. B* **62** 10260 (2000).

21. T. Aono, M. Eto, *Phys. Rev. B.* **63**, 125327 (2001).

22. R. Aguado, D.C. Langreth, *Phys. Rev. B* **67**, 245307 (2003).

23. V. N. Golovach, D. Loss, *Europhys. Lett.* **62**, 83 (2003).

24. Y. Utsumi, J. Martinek, P. Bruno, H. Imamura, cond-mat/0310168 (2003).

25. The device was patterned using surface gates (Cr/Au) fabricated by electron-beam lithography. Peripheral dots have a lithographic area 0.25 $\mu m^2$; the central dot has lithographic area 0.35 $\mu m^2$. The device was fabricated on a delta-doped GaAs/AlGaAs heterostructure with electron gas 100 nm below the surface. Mobility $2x10^5$ cm$^2$/Vs and 2D electron density $2x10^{11}$!cm$^{-2}$ give a transport mean free path of ~2.0 $\mu$m in the unpatterned material.

26. C. M. Varma (personal communication).

27. T. J. Osborne, M. A. Nielsen, *Phys*. *Rev*. *A* **66**, 032110 (2002).

28. A. Saguia, M. S. Sarandy, *Phys*. *Rev*. *A* **67** 012315 (2003).

29. F. Verstraete, M. Popp, J.I. Cirac, *Phys*. *Rev*. *Lett*. **92**, 027901 (2004).

30. We thank Chandra Varma, Bert Halperin, and Amir Yacoby for useful discussion. Research supported in part by the DARPA-QuIST Program, the ARO under DAAD-19-02-1-0070 and DAAD-19-99-1-0215, and the NSF-NSEC Program at Harvard. Research at UCSB was supported in part by iQUIST. NJC acknowledges support by the Harvard HCRP Program. JMT acknowledges support from the NSF. EAL acknowledges support from Middlebury College.

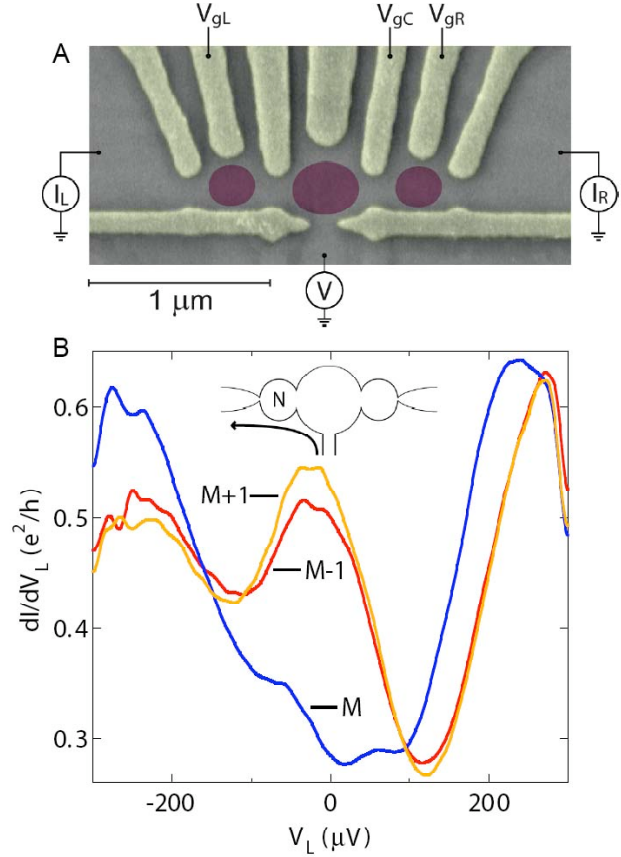

FIG. 1 (A) Scanning electron micrograph of a device identical in design to one measured, with schematic ovals indicating locations of dots upon gate depletion. Gate voltages $V_{gL}$ and $V_{gR}$ change the energies and occupancies of the left and right dots; $V_{gc}$ tunes the coupling of the right dot to the central region. (B) Differential conductance $dI/dV_L$ of the left dot for an odd number of electrons, N. When the right dot contains even number of electrons, (M±1), a zero-bias peak in $dI/dV_L$ is seen, indicating a Kondo state. When the right dot contains an odd number (M) of electrons, the Kondo state in the left dot is suppressed. The states M-1, M, M+1 for the right dot are consecutive Coulomb blockade valleys.

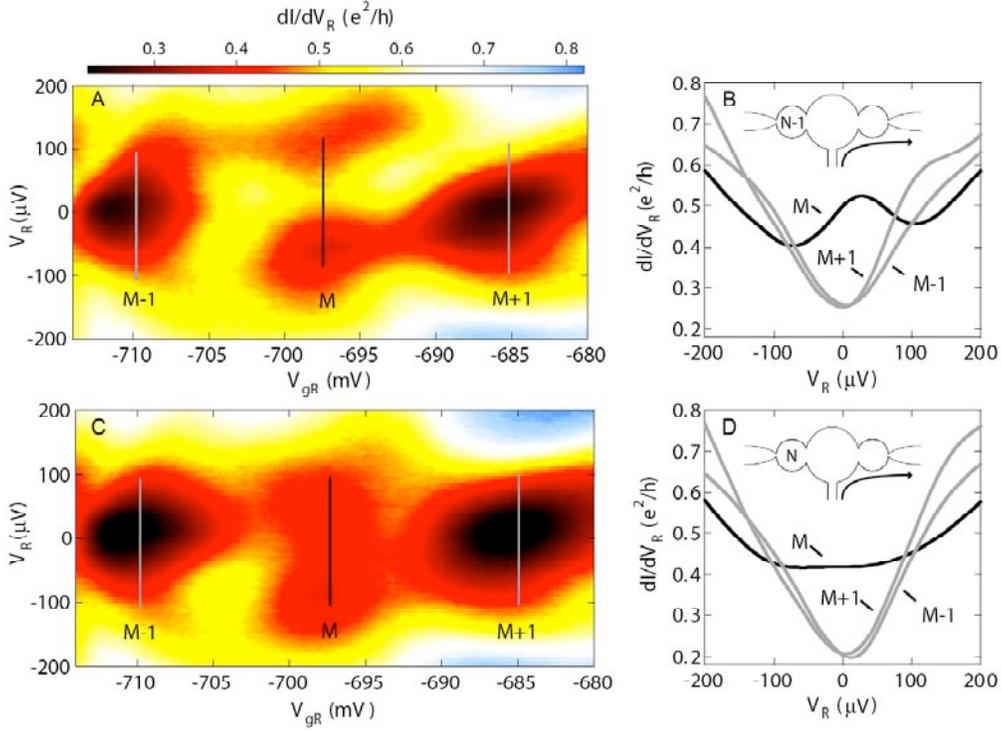

FIG. 2 (A) Differential conductance $dI/dV_R$ of the right dot as a function of both $V_{gR}$ and $V_R$ shows a zero-bias feature for odd occupancy, M. Here, the left dot contains an even number (N-1) of electrons. (B) Slices taken mid-valley from (A) show a zero-bias peak only for odd occupancy M of the right dot. (C) $dI/dV_R$ of the right dot as a function of both $V_{gR}$ and $V_R$, now with an odd number (N) of electrons in the left dot. Suppression of the zero-bias peak in the middle valley is evident. (D) Slices taken mid-valley from (C) show the suppression of the zero-bias peak for the odd-odd (two-impurity) case.

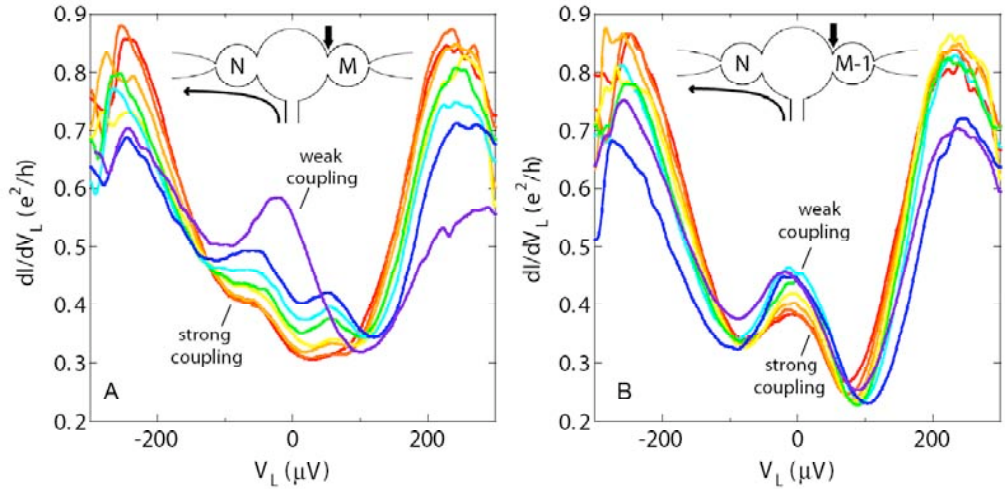

FIG. 3 (A) Differential conductance through the left dot for various values of the coupling between the right dot and the center region. The left dot and right dot both contain odd numbers of electrons (N and M, respectively). For strong couplings, the zero-bias resonance in the left dot is fully suppressed; suppression decreases as the coupling is decreased, so that the zero-bias resonance is fully evident for weak coupling. Notice the splitting of suppressed peaks, which is consistent across a range of couplings. (B) Differential conductance through the left dot for various couplings between the right dot and center region, with an *even* number of electrons (M-1) in the right dot. Traces exhibit a strong zero-bias resonance across all values of the coupling.